# Wavefront distortions of a laser beam reflected from a diffraction grating with imperfect surface and groove pattern

**EFIM KHAZANOV**

*Gaponov-Grekhov Institute of Applied Physics of the Russian Academy of Sciences, 46 Ul'yanov Street, Nizhny Novgorod, 603950, Russia*
*efimkhazanov@gmail.com*

**Abstract:** A formula for the wavefront of a wave reflected from a diffraction grating with an arbitrary surface profile, as well as with arbitrary non-equidistant and non-parallel grooves was obtained. It was shown that the wavefront of the reflected wave can change significantly when the grating is rotated by 180 degrees around the normal. Surface imperfection and groove pattern imperfection are equivalent for monochromatic radiation; whereas for broadband radiation, the two imperfections lead to similar but different space-time coupling effects. For holographic diffraction gratings, wavefront distortions are the sum of distortions caused by grating surface imperfection and the total surface imperfections of the optics used for writing the grating. The second summand is inversely proportional to the frequency of the radiation used to write the grating. The requirements for the flatness of the optics used to write the grating are always more stringent than the requirements for the flatness of the grating.

1. **Introduction**

Diffraction gratings are widely employed in science and technology [1]. The most common use for the diffraction gratings is spectroscopy, where they serve as the wavelength separation device. Spectrometers deliver unique information for the advance in physics, astronomy, chemistry, biology, and medicine. Diffraction gratings are also indispensable in optics and laser physics, where they are used for tuning/narrowing the wavelength of laser generation, for stretching and compressing pico- and femtosecond pulses, for coupling/beamsplitting laser beams, and so on. In this paper, the consideration is restricted to flat reflective gratings, although the proposed description can be generalized to spherical as well as to transmissive gratings. The theoretical description of a diffraction grating is reduced to finding the far field of a wave reflected from the grating if a plane monochromatic wave with frequency $\omega$ is incident on it. If the incident wave is not plane and/or monochromatic, it can be represented as a sum (integral) of plane monochromatic waves, and the reflected field is found using the superposition principle. The reflection coefficient of the grating depends significantly on the groove shape. Optimization of this shape allows achieving high reflection coefficients in a very wide band (see, for example, [2]). In this work, we will not be interested in the amplitude, we will address the phase of the reflected wave, which does not depend on the groove shape.

In the case of an ideal grating, the reflected wave is plane, i.e. its spatial phase $\Delta$ is constant and the angle of reflection $\beta$ is determined by the expression for the grating

$$sin\beta = m\frac{2\pi c}{\omega}\frac{N}{cos\gamma} + sin\alpha, \qquad (1)$$

where $\alpha$ and $\gamma$ are the angles of incidence in the diffraction and orthogonal plane, respectively, $N$ is groove density, and $m$ is the order of diffraction. An ideal grating is understood as a grating with a perfectly flat surface of a substrate and perfectly parallel and equidistant grooves. If the surface is not ideal (non-flat) and the grooves are not ideal (non-equidistant and non-parallel), the wavefront of the reflected wave is no longer flat and $\Delta \neq const$. These wavefront distortions require study. It is important to note that regardless of the cause, these distortions are different for different frequencies, which leads to space-time coupling and entails negative consequences in a number of applications. One of the bright examples is the decrease in the focal intensity of high-power femtosecond lasers. Thus, the description of an imperfect grating reduces to finding the wavefront of the reflected wave, i.e. its spatial phase $\Delta(x, y)$.

The grating non-flatness is caused by inevitable error of substrate polishing, as well as by its thermal loading. The impact of grating non-flatness was numerically studied in many works [3-11] for specific compressor parameters. Different methods of space-time coupling compensation were also discussed in [3-11], but all of them are difficult for experimental implementation. An analytical expression for $\Delta(x, y)$ as well as for the focal intensity of a laser pulse was obtained for gratings of arbitrary surface shape in [12]. It was shown that the decrease in the focal intensity depends on the product of grating surface rms and pulse spectrum bandwidth. With low-quality gratings, spectrum narrowing does not reduce focal intensity; contrariwise, it may even slightly increase it.

The influence of grating groove imperfections on the spectrometer resolution was investigated in [13-15] and on the operation of a femtosecond laser pulse compressor in [16]. However, these works did not give a solution to the problem in a general form: the expression for the spatial phase of the reflected wave $\Delta(x, y)$ was not derived. Grating groove imperfections (in contrast to its non-flatness) are very hard to measure in practice [17]. For holographic gratings, groove imperfections are determined exclusively by the imperfection of the wavefronts of the waves used for writing the grating [1, 16]. The influence of these fronts was investigated in [18], and a method for their minimization was proposed in [19]. However, these studies were limited to Gaussian beams, i.e. to beams with a parabolic wavefront containing only one Zernike polynomial (defocus). In practice, distortions are inevitably much more complex and are described by a significantly larger number of polynomials. Thus, imperfection of the groove pattern has been studied much less than grating non-flatness, so comparison of the influence of these two effects on the grating quality has been impossible until now.

The paper is organized as follows. In the second section, an analytical formula is derived for the spatial phase $\Delta(x, y)$ of the wave reflected from the grating for an arbitrary profile of its surface and arbitrary imperfection of its grooves. In the third section, this phase is expressed through the grating surface profile and the total profile of the surfaces of the optical elements used to write the grating. The following two sections provide examples of using the obtained expressions: the degradation of the Czerny-Turner spectrometer resolution is found in section 4 and the focal intensity decrease after laser pulse compression in the Treacy compressor in section 5.

## 2. General view of the expression for the phase of the beam reflected from the diffraction grating

Let a plane monochromatic wave with frequency $\omega$ be incident on a diffraction grating at an angle $\alpha$ in the $x''z''$ plane and at an angle $\gamma$ in the $y''z''$ plane (Fig. 1a). If the grating is ideal, the reflection angle $\beta$ is determined by formula (1). Hereinafter we count the angle of incidence counterclockwise ($\alpha$>0 in Fig. 1a) and the reflection angle clockwise ($\beta$<0 in Fig. 1a). Let us consider an imperfect grating, the surface profile of which is described by the function $h_{gr}(x'', y'')$. In addition, the groove pattern is also imperfect (Fig. 1b): $N \neq const$ and the grooves may be nonparallel to the $y''$ axis, i.e. the groove pattern is described by the vector $\boldsymbol{K}(x'', y'')$, whose components $K_{x,y}(x'', y'')$ correspond to the local groove density along the $x''$ and $y''$ coordinates. At each $(x'', y'')$ point, the vector $\boldsymbol{K}$ is perpendicular to the grooves, and the distance between the grooves in this direction is $1/|\boldsymbol{K}|$. We will assume that the imperfection of the grooves is small:

$$\boldsymbol{K}(x'', y'') = N \begin{pmatrix} 1 + \delta(x'', y'') \\ \chi(x'', y'') \end{pmatrix}, \qquad (2)$$

where $\delta \ll 1$ means that the grooves are nonequidistant, and $\chi \ll 1$ is the angle of incidence of the grooves to the $y$ axis. For an ideal grating, $\delta(x'', y'') = \chi(x'', y'') = h_{gr}(x'', y'') = 0$. The phase difference between the incident and reflected waves $\Delta(x', y')$ is equal to the phase difference between the points A and C and is determined by the expression [14, 20]:

$$\Delta(x', y') = k_{zx}|AB| + k_{zx}|BC| + m2\pi M_{OD}, \quad (3)$$

where $M_{OD}$ is the number of grooves between the points O and D, and $k_{zx} = \frac{\omega}{c}\cos\gamma$. Strictly speaking, the angle of reflection from an imperfect grating is not equal to $\beta$ (the angle of incidence is different from $\alpha$ and the groove density is not equal to $N$); therefore, the distance from the point B to the $Ox'$ axis will be larger than $|BC|$, but this elongation is negligible. The expression for $M_{OD}$ has the form [14]:

$$M(x'', y'') = \int_0^{x''+|FD|}(N + N\delta(x_1, 0))dx_1 + \int_0^{y''} N\chi(x'', y_1)dy_1 \quad (4)$$

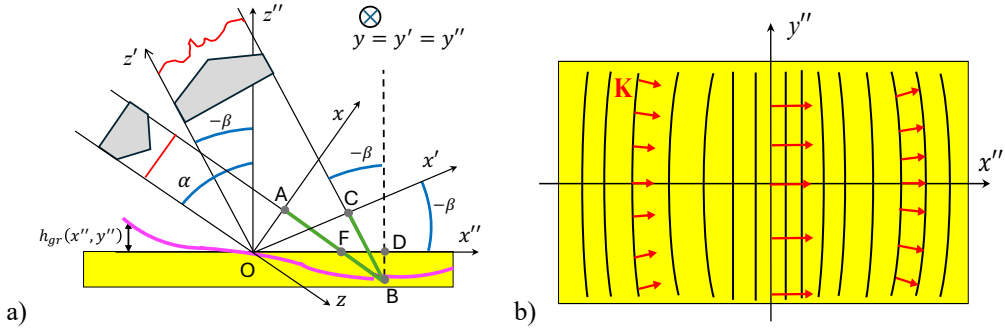

Fig. 1. (a) Wavefront distortions at reflection from diffraction grating. The beam incident on the grating is written in the $(x, y)$ coordinate system, reflected in the $(x', y')$ coordinate system, and the grating plane is in the $(x'', y'')$ coordinate system. (b) Groove pattern, the red arrows show the vector $\boldsymbol{K}(x'', y'')$.

The substitution of (4) into (3), after transformations yields

$$\Delta(x', y', \omega) = -\frac{\omega}{c}\cos\gamma(\cos\alpha + \cos\beta)h\left(\frac{x'}{\cos\beta}, \frac{y'}{\cos\gamma}\right) + 2\pi mNg\left(\frac{x'}{\cos\beta}, \frac{y'}{\cos\gamma}\right) \quad (5a)$$

or, taking into account (1), an equivalent expression

$$\Delta(x', y', \omega) = -\frac{\omega}{c}\cos\gamma(\cos\alpha + \cos\beta)\left\{h\left(\frac{x'}{\cos\beta}, \frac{y'}{\cos\gamma}\right) + tg\left(\frac{\alpha-\beta}{2}\right)g\left(\frac{x'}{\cos\beta}, \frac{y'}{\cos\gamma}\right)\right\}, \quad (5b)$$

where

$$g(x'', y'') = \int_0^{x''}\delta(x_1, 0)dx_1 + \int_0^{y''}\chi(x'', y_1)dy_1 \quad (6)$$

The function $g(x'', y'')$ characterizes the imperfection of the grooves at the point $(x'', y'')$ on the grating surface. In the case of ideal grooves, $g = 0$ and (5) transforms to the formula obtained in [12] in a different way. A non-trivial consequence of (5, 6) is a pronounced change in the phase of the wave reflected from the grating $\Delta(x', y', \omega)$ when it is rotated by 180 degrees around the $z''$ axis. The fact is that with such a rotation the function $h(x'', y'')$ transforms to $h(-x'', -y'')$ and the function $g(x'', y'')$, as seen from (6), transforms to $-g(-x'', -y'')$. Thus, in front of the term with $g$ in (5) we can put the sign $\pm$ instead of plus, and for minimizing phase distortions we can choose any sign.

The wavefront distortions $\Delta$ at reflection from an imperfect grating are the sum of the distortions caused by the imperfect surface $h$ and the imperfect groove pattern $g$. As can be seen from (5a), the contribution of the last term to $\Delta$ is independent of $\omega$, which creates a deceptive impression of the absence of space-time-coupling – the phase distortions are the same for all frequencies. Actually, there is no space-time-coupling if the optical path $c\Delta/\omega$ is independent of $\omega$. As can be seen from (5a), the optical path is proportional to $\omega^{-1}$. For monochromatic radiation, it follows from (5) that the surface imperfection is equivalent to the groove pattern imperfection, since the latter can be interpreted as an additive to the former and vice versa. For example, the curly bracket in (5b) can be considered as the effective grating surface profile $h_{eff}$. It is important to note that this is not the case for broadband radiation, since the expression in curly bracket depends on frequency and, therefore, cannot be interpreted as a surface profile. In the next section we will derive the expression for $g(x,y)$ for holographic diffraction gratings.

## 3. Relationship between the $g(x,y)$ function and the phase of the beams writing the diffraction grating

The form of $g(x,y)$ can be defined more concretely for holographic diffraction gratings written using the interference of two waves (see Fig. 2). In this case, the reason for imperfect density of the grooves is the distortion of the wavefront of these waves [1, 16]. Let the grating be written by two waves with frequency $\omega_{wr} = ck_{wr} = 2\pi c/\lambda_{wr}$ incident at an angle $\Phi$ to the normal of the grating (substrate). The angle $\Phi$ is defined by the apparent relation

$$\sin\Phi = \frac{N\lambda_{wr}}{2} \tag{7}$$

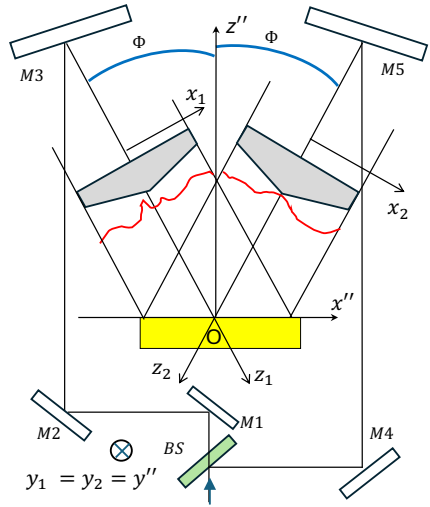

Fig. 2. Scheme of holographic diffraction grating writing by two laser beams. The writing beams are described in $(x_1, y_1)$ and $(x_2, y_2)$ coordinates and the grating plane in - $(x'', y'')$ coordinates.

The case of Gaussian writing beams with a parabolic wavefront was considered in [18, 19]. We will obtain a solution in a general form, when both writing waves have arbitrary transverse phases $\psi_1$ and $\psi_2$. The fields of these waves in the reference frames with indices 1 and 2 have the form

$$E_{1,2}(z_{1,2}, \boldsymbol{r}_{1,2}) = \mathcal{E} e^{ik_{wr}z_{1,2}} e^{i\psi_{1,2}(\boldsymbol{r}_{1,2})} = \mathcal{E} e^{ik_{wr}z_{1,2}} e^{i\boldsymbol{\kappa}_{1,2}(\boldsymbol{r}_{1,2})\boldsymbol{r}_{1,2}}, \tag{8}$$

where $\boldsymbol{\kappa}_{1,2}(\boldsymbol{r}_{1,2}) = \nabla\psi_{1,2}(\boldsymbol{r}_{1,2})$. Multiplication of $\boldsymbol{\kappa}_{1,2}(\boldsymbol{r}_{1,2})$ by the matrix of rotation by the angle $\pm\Phi$ enables finding their transverse wave vectors $\boldsymbol{\kappa}_{1,2}(\boldsymbol{r}'')$ on the grating surface and, next, the vector $\boldsymbol{K}$ from the expression

$$\boldsymbol{K}(\boldsymbol{r}'') = \frac{\boldsymbol{\kappa}_1(\boldsymbol{r}'') - \boldsymbol{\kappa}_2(\boldsymbol{r}'')}{2\pi} \tag{9}$$

The substitution of (6) and (9) into (2) gives

$$g(x,y) = \frac{h_{wr}(x\cos\Phi, y)}{\sin\Phi}, \tag{10}$$

where $h_{wr}(x,y) = \frac{\psi_1(x,y) - \psi_2(x,y)}{2k_{wr}}$ has the meaning of the difference in the total surface profiles of the optical elements on the path of two waves between the splitter and the substrate. Transmissive optics (for example, a beamsplitter) can be recalculated to an equivalent surface. In (10), without loss of generality, we set $h_{wr}(0,0) = 0$. Substituting (10) into (5b) yields

$$\Delta(x', y', \omega) = -\frac{\omega}{c}\left(H_{gr} h_{gr}\left(\frac{x'}{\cos\beta}, \frac{y'}{\cos\gamma}\right) + H_{wr} h_{wr}\left(\frac{x'\cos\Phi}{\cos\beta}, \frac{y'}{\cos\gamma}\right)\right), \tag{11}$$

where

$$H_{gr} = \cos\gamma(\cos\alpha(\omega) + \cos\beta(\omega)) \qquad H_{wr} = 2\frac{\omega_{wr}}{\omega} \tag{12}$$

Finally, we obtain the relationship between the incident and reflected fields:

$$E_{ref}(x', y') = \sqrt{\frac{\cos\beta}{\cos\alpha}} E_{in}\left(\frac{\cos\beta}{\cos\alpha} x', y'\right) \exp(i\Delta(x', y', \omega)) \tag{13}$$

Thus, the wavefront distortions of the field reflected from the grating, caused by the imperfection of the grating grooves, are equal to the algebraic sum of the surface profiles of all elements of the transport optics $h_{wr}$ to an accuracy of up to the factor $H_{wr}$. Note that from the point of view of reducing the inhomogeneity of the grooves (reduction of $H_{wr}$), it is better to write the grating by radiation with a lower frequency $\omega_{wr}$. This has a simple physical explanation: for a fixed $h_{wr}$, the smaller $\omega_{wr}$, the smaller the introduced phase distortions.

Using (11), it is convenient to compare the requirements for the accuracy of the grating surface $h_{gr}(x,y)$ with the requirements for the accuracy of the surfaces of the optical elements used for its writing, $h_{wr}(x,y)$. For example, for $\alpha = 53°$, $\gamma = 0°$, $N = 1200/\text{mm}$, $\lambda_{wr} = 413\text{nm}$, and $\lambda_0 = 800\text{nm}$ we obtain $H_{wr} = 2.5 H_{gr}$, i.e. the requirements for the quality of the optics used to write the grating are significantly higher than the requirements for the quality of the grating substrate. Note also that $h_{gr}(x,y)$ is the profile of one surface, and $h_{wr}(x,y)$ is the total profile of all surfaces. In Fig. 2, they are a beamsplitter and mirrors M1-M5.

In the next two sections, using (11, 13) we consider the impact of grating imperfection on the operation of the spectrometer and the compressor of femtosecond laser pulses.

## 4. The influence of grating imperfection on the resolution of the Czerny-Turner spectrometer

Let us consider as an example the Czerny-Turner spectrometer, in which frequency scanning is performed by rotating the grating (Fig. 3). For this spectrometer, both the angle of incidence $\alpha$ and the angle of reflection $\beta$ are frequency dependent, but their sum does not depend on frequency: $\alpha(\omega) + \beta(\omega) = 2\phi = const$. The spectrometer resolution is determined by the monochromatic beam spreading at the output slit in the direction of the $x'$ axis. We will assume that aberration-free off-axis parabolas with focus $F$ are used as mirrors M1 and M2. Then the beam size at the slit is determined by the diffraction limit and the imperfection of the grating. The field reflected from mirror M2 has the form

$$E_2(x', y') = E_{ref}(x', y') e^{i\frac{kr'^2}{2F}} = \sqrt{\frac{\cos\alpha}{\cos\beta}} E_{in}\left(\frac{\cos\alpha}{\cos\beta} x', y'\right) \exp(i\Delta(x', y', \omega)) e^{i\frac{kr'^2}{2F}} \tag{14}$$

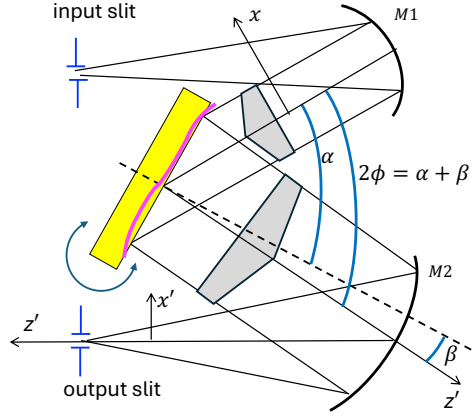

Fig. 3. Czerny-Turner spectrometer scheme. The beam incident on the grating is described in $(x, y)$ coordinates and the reflected beam in $(x', y')$ coordinates. M1, M2 are mirrors.

Here we neglected the diffraction between the grating and mirror M2. The second moment of the beam $X$ in the plane of the output slit is used as the resolution criterion

$$X^2 = \frac{\int x'^2 |E_{slit}(x',y')|^2 dx'dy'}{\int |E_{slit}(x',y')|^2 dx'dy'},$$

where $E_{slit}(x', y')$ is the field in the plane of the output slit, which is located in the plane of the beam waist. For an ideal grating ($\Delta = 0$), this plane is the same for all frequencies and the resolution is determined by the diffraction limit $X(\Delta = 0) = X_{diff}(\omega)$. For an imperfect grating, due to space-time coupling effects, the waist position depends on frequency. Obviously, the optimal choice is to locate the slit in the waist plane for the central frequency $\omega_0$. We designate the distance from this waist to mirror M2 as $z$ dependent on the phase $\Delta(x', y', \omega_0)$. To find $X$ we use the moments method [21], according to which

$$X^2 = W^2 + Bz + Cz^2, \tag{15}$$

where

$$W^2 = \frac{\int x'^2 A^2 dx'dy'}{\int A^2 dx'dy'} \quad B = \frac{2}{k} \frac{\int x' \frac{\partial \varphi}{\partial x'} A^2 dx'dy'}{\int A^2 dx'dy'} \quad C = \frac{1}{k^2} \frac{\int \left(\frac{\partial A}{\partial x'}\right)^2 dx'dy'}{\int A^2 dx'dy'} + \frac{1}{k^2} \frac{\int \left(\frac{\partial \varphi}{\partial x'}\right)^2 A^2 dx'dy'}{\int A^2 dx'dy'}, \tag{16}$$

$A(x', y')$ and $\varphi(x', y')$ are the amplitude and phase of the field (14). Let a Gaussian beam

$$E_{in}(x, y) = e^{-\frac{x^2 + y^2}{2w^2}} \tag{17}$$

be incident on the grating.

Consider three types of surface distortions of the optical elements $h_{wr}(x, y)$ used for writing: defocus, vertical astigmatism, and oblique astigmatism. For the resolution $R = X/X_{diff}(\omega_0)$ normalized to the diffraction limit at central frequency, from (14, 15, 16, 17) we obtain the following expressions

$$R_{wr,d}^2 = \rho^2 + 2\left(\frac{\omega_0 h_{pv}}{c} \frac{\cos\beta(\omega_0)}{\cos\alpha(\omega_0)} \frac{\cos^2\Phi}{\cos^2\beta(\omega)} \left(H_{wr}(\omega_0) - H_{wr}(\omega)\right)\right)^2 \quad \text{for defocus} \tag{18a}$$

$$R_{wr,va}^2 = \rho^2 + \frac{1}{2}\left(\frac{\omega_0 h_{pv}}{c} \frac{\cos\beta(\omega_0)}{\cos\alpha(\omega_0)} \frac{\cos^2\Phi}{\cos^2\beta(\omega)} \left(H_{wr}(\omega_0) - H_{wr}(\omega)\right)\right)^2 \quad \text{vertical astigmatism} \tag{18b}$$

$$R^2_{wr,oa} = \rho^2 + \frac{1}{8}\left(\frac{\omega_0 h_{pv}}{c}\frac{\cos\beta(\omega_0)}{\cos\alpha(\omega_0)}\frac{\cos\Phi}{\cos\beta(\omega)\cos\gamma}H_{wr}(\omega)\right)^2 \qquad \text{for oblique astigmatism,} \quad (18c)$$

where $h_{pv}$ is $h_{wr}(x, y)$ peak-to-valley (PV) for a circle of radius $w$ and

$$\rho = \frac{\omega_0}{\omega}\frac{\cos\alpha(\omega)}{\cos\beta(\omega)}\frac{\cos\beta(\omega_0)}{\cos\alpha(\omega_0)}$$

To find $R^2_{gr}$ for similar distortions of the grating surface, it is necessary to make the following substitutions in (18): $H_{wr} \to H_{gr}$; $\cos\Phi \to 1$. Figure 4 shows $R(\lambda)$ for $\lambda_{wr} = 413$nm, $2\phi = 40°$, $N = 1400$/mm, and $h_{pv} = \lambda_0$. As expected, defocus and vertical astigmatism are completely compensated for only at the central frequency $R(\lambda = 900\text{nm}) = 1$ and for oblique astigmatism even $R(\lambda = 900\text{nm}) > 1$. It is easy to show that this is also true for higher-order aberrations (coma, spherical aberration, etc.). For almost all parameters $R_{wr} \gg R_{gr}$, i.e. the greatest influence on the resolution is exerted by the aberrations of the optics used for writing the grating, rather than by the inaccuracy of manufacturing the grating surface.

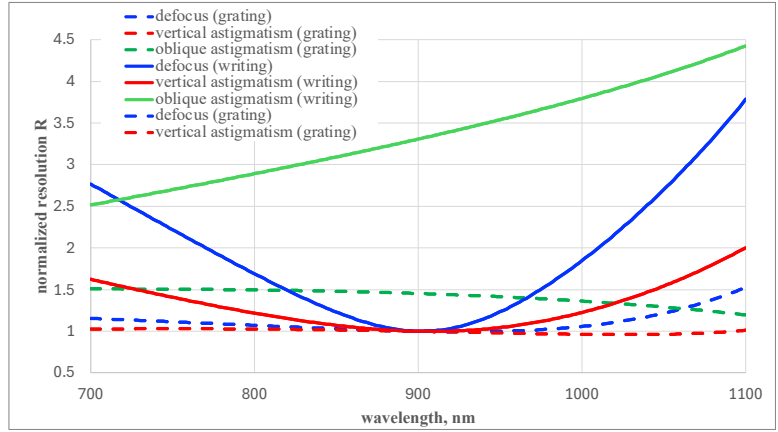

Fig. 4. Normalized spectrometer resolution versus wavelength for different types of aberrations: defocus (blue), vertical astigmatism (red), oblique astigmatism (green); solid curves are aberrations of the optics used to write the grating, dashed lines are aberrations of the grating. The peak-to-valley for all curves is $h_{pv} = \lambda_0$.

## 5. The influence of grating imperfection on the radiation focal intensity after compressor

In recent years, in addition to the classical four-grating Treacy compressor [22] (Fig. 5), other types of compressors have been actively studied for high-power femtosecond lasers. These include an asymmetric compressor [23-25], an out-of-plane compressor [26-29], a compressor with two gratings [30-32], and compressors with six gratings [33, 34]. An analytical expression for the focal intensity of a laser pulse depending on the surface shape of diffraction gratings $h_{gr}(x, y)$ for an arbitrary compressor was obtained in a general form in [12]. It was shown that in the Treacy compressor the impact of grating non-flatness $h_{gr}(x, y)$ on the focal intensity is due to two effects [12]. The first one is the spatial chirp of the beam on gratings G2 and G3 as a result of which different frequencies "see" different surface shapes on reflection from these gratings. The second effect is that when reflected from the grating, the phase front repeats the shape of the grating, but the proportionality coefficient depends on frequency, see (5). This effect is present in all four gratings.

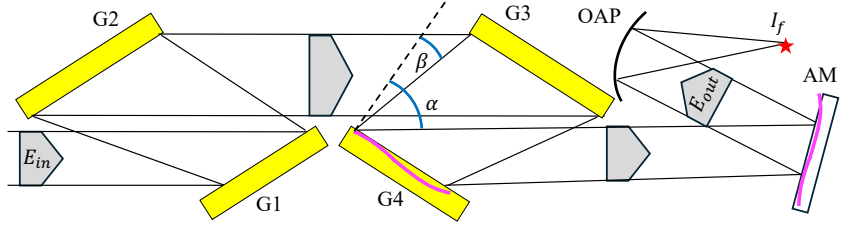

Fig. 5. Beam focusing after Treacy compressor. G1-G4 – diffraction gratings, AM – adaptive mirror, OAP – off-axis parabola.

A full generalization of the results of [12] to the groove imperfection $h_{wr}(x,y)$, as well as a detailed comparison of the influence of the surface and groove imperfection will be addressed in a separate publication. Here we will restrict consideration to the simplest case of a symmetric four-grating Treacy compressor, in which only the last grating is not ideal (Fig. 5). We have retained the traditional notation of angles for compressors: $\alpha, \beta$ are the angles of incidence and reflection from the first grating and vice versa for the fourth grating. The field incident on the compressor in the $(r, \omega)$-space will be represented as a beam of an arbitrary profile with a Gaussian spectrum of width $\Delta\omega$:

$$E_{in}(x,y,\omega) = E_0(x,y)e^{i\phi_{in}(x,y)}e^{-\frac{(\omega-\omega_0)^2}{2(\Delta\omega)^2}}e^{i\varphi_{in}(\omega)}, \tag{19}$$

where $\varphi_{in}(\omega)$ is the spectral phase introduced by all optics before the compressor: generator, stretcher, amplifier, and acousto-optic programmable dispersive filter [35]. Then, can be readily shown that the field after the adaptive mirror will have the form:

$$E_{out}(x,y,\omega) = E_{in}(x,y,\omega)e^{i\varphi_{am}(x,y)}e^{i\tilde{\Delta}(x,y,\omega)}e^{i\varphi_c(\omega)}, \tag{20}$$

where

$$\tilde{\Delta}(x,y,\omega) = -\frac{\omega}{c}\left(H_{gr}(\omega)h_{gr}\left(\frac{x}{\cos\alpha},\frac{y}{\cos\gamma}\right) + H_{wr}(\omega)h_{wr}\left(\frac{x\cos\Phi}{\cos\alpha},\frac{y}{\cos\gamma}\right)\right), \tag{21}$$

$\varphi_{am}(x,y)$ is the phase introduced by an adaptive mirror and $\varphi_c(\omega)$ is the phase introduced by a compressor with ideal gratings. The expression for $\tilde{\Delta}(x,y,\omega)$ (21) slightly differs from (11), as for the fourth grating the angle of reflection is $\alpha$, rather than $\beta$. Note that, in contrast to (11), $h_{gr}$ and $h_{wr}$ are frequency independent, since $\alpha$ does not depend on frequency. It will be further assumed that the compressor dispersion matches the input spectral phase, i.e. $\varphi_{in}(\omega) = -\varphi_c(\omega)$, and the adaptive mirror compensates for the distortions of the input beam wavefront $\phi_{in}(x,y)$ and the distortions $\tilde{\Delta}(x,y,\omega_0)$ introduced by the fourth grating for the central frequency $\omega_0$, i.e. $\varphi_{am}(x,y) = -\tilde{\Delta}(x,y,\omega_0) - \phi_{in}(x,y)$.

The field in the focal plane is proportional to the three-dimensional Fourier spectrum of $E_{out}$ that will be denoted as $E_{out}(\kappa_x, \kappa_y, t)$. The maximum field is $E_{out}(\kappa_x=0, \kappa_y=0, t=0)$; hence, the focal intensity $I_f$ will be

$$I_f \sim |\int E_{out}(x,y,\omega)dxdyd\omega|^2, \tag{22}$$

and the Strehl ratio $St$ characterizing the decrease in $I_f$ due to grating imperfection will be

$$St = \frac{I_f}{I_f(h_{gr}=h_{wr}=0)}. \tag{23}$$

In the $\Delta\omega \ll \omega_0$ approximation, from (20, 22, 23) we find

$$St \approx \left( \frac{\int E_0(x,y)dxdy \exp\left(-\left(\frac{\Delta\omega}{\omega_0}\right)^2 \frac{(2\pi N\Psi(x,y))^2}{2}\right)}{\int E_0(x,y)dxdy} \right)^2, \quad (24)$$

where

$$\Psi = tg\beta(\omega_0) h_{gr}\left(\frac{x}{\cos\alpha}, \frac{y}{\cos\gamma}\right) + \frac{1}{\sin\Phi} h_{wr}\left(x\frac{\cos\Phi}{\cos\alpha}, \frac{y}{\cos\gamma}\right) + const, \quad (25)$$

and $const$ is chosen from the condition of $St$ maximization. Note that $St$ is determined by $h_{gr,wr}$, rather than by $\frac{\partial h_{gr,wr}}{\partial x}$, as the resolution of the spectrometer (see (15, 16)). Similarly to the previous section, we will consider three types of $h_{wr}(x,y)$: defocus, vertical astigmatism, and oblique astigmatism. The field $E_0(x,y)$ will be regarded to be Gaussian (17). Then, under the condition of small phase distortions $\frac{\Delta\omega}{\omega_0}\pi N\Psi(x,y) \ll 1$, the exponent in (24) can be expanded into a Taylor series and after integration the Strehl ratio will be obtained:

$$St_{wr} \approx 1 - \frac{1}{2}\left(\pi N \frac{\Delta\omega}{\omega_0} \frac{h_{pv}}{\sin\Phi\cos^2\alpha}\right)^2 (\cos^4\Phi + \cos^4\alpha) \quad \text{for defocus} \quad (26a)$$

$$St_{wr} \approx 1 - \frac{1}{2}\left(\frac{\pi N}{2} \frac{\Delta\omega}{\omega_0} \frac{h_{pv}}{\sin\Phi\cos^2\alpha}\right)^2 (\cos^4\Phi + \cos^4\alpha) \quad \text{for vertical astigmatism} \quad (26b)$$

$$St_{wr} \approx 1 - \left(\frac{\pi N}{2} \frac{\Delta\omega}{\omega_0} \frac{\cot\Phi}{\cos\alpha} h_{pv}\right)^2 \quad \text{for oblique astigmatism} \quad (26c)$$

The expressions for $St_{gr}$ coincide with (26), if $\sin\Phi, \cos\Phi$, and $\cot\Phi$ are replaced by 1 and $h_{pv}$ is replaced by $h_{pv} tg\beta(\omega_0)$. The curves for $St(h_{pv})$ are plotted in Fig. 6 for the parameters typical of high-power femtosecond lasers: $\lambda_0 = 800$nm, $\Delta\omega = 0.08\omega_0$, $N = 1400$/mm, $\alpha = 45°$, and $\lambda_{wr} = 413$nm.

As expected, the influence of the aberrations of the optics used for writing the grating is much greater than the manufacturing inaccuracy of the grating itself. It follows from Fig. 6 that the requirements for the quality of the optics seem to be quite feasible in practice: if the combined profile of the beamsplitter and mirrors M1-M5 (Fig. 2) has a peak-to-valley $\lambda/5$, the Strehl ratio $St > 0.85$. It is important to note that these results were obtained for an imperfect fourth grating, while the others were considered ideal (Fig. 5). The same result will be for an imperfect first grating. However, as was shown in [12], the influence of the second and third gratings is much stronger, since the spatial chirp of the beam on these gratings, due to which different frequencies are reflected from different grating profiles, leads to a much stronger decrease in the focal intensity. A detailed study of this issue will be presented in a separate publication.

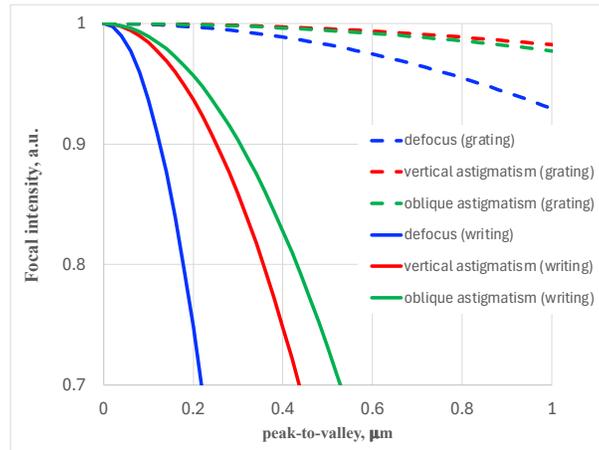

Fig. 6. Focal intensity versus aberration of the fourth compressor grating: defocus (blue), vertical astigmatism (red), oblique astigmatism (green); solid curves – aberrations of the optics used to write the grating, dashed curves – aberrations of the grating.

## 6. Conclusion

The wavefront distortions $\Delta$ of the wave reflected from the diffraction grating are the sum of the distortions caused by the surface imperfection $h_{gr}(x,y)$ and the groove imperfection $g(x,y)$ and are described by the obtained expression (5). The coefficients before these terms depend on frequency, so for non-monochromatic radiation these two effects, despite their similarity, are not equivalent and cannot be reduced to each other. The value of $\Delta$ changes significantly when the grating is rotated by 180 degrees around the normal. In other words, there are two variants to use the same grating. The variant when the distortions lead to minimal negative consequences should be chosen in practice.

For holographic diffraction gratings $g(x,y)$ is equal, to an accuracy of the multiplier, to the total profile $h_{wr}(x,y)$ of the surface of all elements of the transport optics used for writing the grating, and the wavefront distortions $\Delta$ are determined by (11). Thus, $\Delta$ is the sum of the distortions caused by the imperfection of the surface of the grating $h_{gr}(x,y)$ and the total imperfection of the surfaces of the optics used to write it $h_{wr}(x,y)$. For identical values of $h_{gr}(x,y)$ and $h_{wr}(x,y)$, the latter has a greater negative effect on both the spectrometer (Fig. 4) and the femtosecond pulse compressor (Fig. 6). This imposes much more stringent requirements for the flatness of the optics used for writing the grating, compared to the requirements for the flatness of the grating itself. From the point of view of reducing the inhomogeneity of the grooves, it is better to write the grating by radiation with a lower frequency.

**Funding.** Ministry of Science and Higher Education of the Russian Federation (075-15-2020-906; Center of Excellence "Center of Photonics").

**Disclosures.** The authors declare no conflicts of interest.

**Data availability.** Data underlying the results presented in this paper are not publicly available at this time but may be obtained from the authors upon reasonable request.